# Nuclear Quantum Effects on Proton Diffusivity in Perovskite Oxides

Shunya Yamada, Kansei Kanayama, and Kazuaki Toyoura*

*Department of Materials Science and Engineering, Kyoto University, Kyoto 606-8501, Japan*

* toyoura.kazuaki.5r@kyoto-u.ac.jp

**Abstract**

In the present study, the nuclear quantum effects (NQEs) on proton diffusivity in oxides were evaluated by molecular dynamics (MD) simulations with the quantum thermal bath (QTB) based on the Langevin dynamics. We employed the proton diffusion in barium zirconate ($BaZrO_3$) with the cubic perovskite structure as the model system, in which protons migrate by *rotation* around single oxide ions and *hopping* between adjacent oxide ions. MD simulations with the standard classical thermal bath (CTB) and phonon calculations were also conducted to verify the conventionally used classical harmonic transition state theory (classical h-TST), in which the transition state theory (TST), the harmonic approximation, and the classical approximation are assumed. As a result, the h-TST are reasonable for the proton rotation, while significantly overestimate the activation energy and the pre-exponential factor of the jump frequency for the proton hopping. Furthermore, the classical approximation makes the proton jump frequencies close to linear in the Arrhenius plots, which should actually be nonlinear by the NQEs in the temperature range of 500–2000 K. This suggests the necessity of the treatment beyond the classical h-TST for accurate evaluation of the proton diffusivity in oxides even in the intermediate temperature range (573–873 K).

# I. Introduction

Proton-conducting oxides are expected to be applied to electrolytes in various electrochemical devices, such as fuel cells, water electrolyzers, and gas sensors. Many proton-conducting oxides with various crystal structures have been reported so far [1–6] in which acceptor-doped barium zirconate and cerate ($BaZrO_3$ and $BaCeO_3$) with the perovskite-type structure are promising candidates for the electrolytes with high proton conductivity beyond $10^{-2}$ S/cm at 873 K [7–10]. However, because of their practical problems, i.e., hole conductivity under oxidizing atmosphere, chemical instability to carbon dioxide, and poor sinterability [11,12] novel classes of proton-conducting oxides have still been explored [13].

The fundamental understanding of the proton conduction mechanism is of importance in terms of materials exploration for high-performance electrolytes. Theoretical calculations, particularly first-principles calculations, are powerful tools to reveal the atomic-scale picture of proton conduction mechanism in oxides, i.e., *rotation & hopping* [2,14,15] and to quantitatively evaluate the proton-acceptor and the proton-proton interactions [16–19]. The computational approaches for investigating ionic conduction are mainly classified into two types, the approaches based on the equation of motion and the statistical-mechanics-based approaches. The former is well known as molecular dynamics (MD) simulations, in which the atomic motion is conventionally treated within classical mechanics. MD simulations provide information on the time evolution of a given system, and we can obtain the trajectory of every atom in the system including mobile protons. In addition, the proton diffusivity and mobility can be evaluated quantitatively from the mean square displacement (MSD). The only critical drawback of MD simulations is *time-scale limitation*, e.g., sub-nano-second time scale ($< 10^{-9}$ s) at most in the case of first-principles MD (FPMD) simulations.

The statistical-mechanics-based approaches can overcome the time-scale limitation, which requires the potential energy surface (PES) in the configuration space. Based on the transition state theory (TST) [20], the atomic jump frequencies, $\Gamma$, are approximately given by the Arrhenius-type equation,

$$\Gamma = \Gamma_0 \exp\left(-\frac{\Delta U}{k_B T}\right), \qquad (1)$$

where $\Gamma_0$, $\Delta U$, $k_B$, and $T$ are the vibrational prefactor, the potential energy barrier for an atomic jump, the Boltzmann constant, and the temperature, respectively. $\Delta U$ is generally evaluated by the nudged elastic band (NEB) method [21]. In the case of proton jumps in oxides, all proton sites are first explored exhaustively, and the potential barriers for all possible migration paths between adjacent proton sites are then evaluated by the NEB method. $\Gamma_0$ is often approximated by a constant value comparable to lattice vibration, e.g. $10^{13}$ /s or estimated from the eigen-frequencies at the initial and saddle-point states for a proton jump under the classical harmonic approximation [20]. The proton diffusivity and mobility are finally estimated using all proton jumps through the kinetic Monte Carlo simulations [15,22,23] or the master-equation approach [17].

However, both conventional methods based on classical mechanics are considered to be insufficient for proton dynamics, in which the nuclear quantum effects (NQEs) are not negligible. The importance of NQEs was already recognized for hydrogen dynamics in metallic systems [24] while the NQEs for proton dynamics in oxides has often been neglected because the temperature range of interest is relatively high, 573–873 K. Zhang et al. addressed the NQEs for proton dynamics in $BaZrO_3$ by the path integral molecular dynamics (PIMD) [25], suggesting that the NQEs are not negligible below 600 K. They pointed out that proton rotation (OH reorientation) is the rate-limiting step below 600 K in the PIMD, in contrast to the rate-limiting proton hopping (proton transfer) in conventional MD simulations. Recently, Geneste quantitatively evaluated the adiabatic and non-adiabatic proton tunneling for the proton rotation and hopping in perovskite oxides [26–28], where the proton jump rates were evaluated as a function of temperature in the range of 150–900 K.

In the present study, we have revisited the NQEs in proton dynamics in oxides based on Langevin MD with the quantum thermal bath (QTB-MD) [29–32]. In the QTB-MD, the NQEs are incorporated by using the time-correlated random forces instead of the time-uncorrelated ones in the conventional Langevin thermostat, while each atom in a given system is still treated as a classical particle. The proton diffusivities can easily be estimated from the counted jump frequencies of proton rotation and hopping during the simulations, where the quantum transition state theory employed in the literature [25] is not assumed. In addition, this method does not assume the Born-Oppenheimer approximation in terms of the proton state, which is often utilized in the previous studies on the NQEs of hydrogen atoms or protons [26–28]. The on-the-fly machine learning force field molecular dynamics (MLFF-MD) simulations combined with first-principles electronic structure calculations [33–35] were here employed for reducing the huge computational costs of first-principles QTB-MD simulations. The obtained results are compared with the results estimated within the framework of the harmonic transition state theory (h-TST) to estimate the anharmonicity and to verify the TST, for which the eigen-frequencies of lattice vibrations were estimated at both initial and saddle-point states for proton rotation and hopping. In addition, the conventional Langevin MD with the classical thermal bath (CTB) were also performed for comparison. The proton dynamics in the perfect crystal of $BaZrO_3$ with the cubic perovskite structure was taken as a model system in the present study.

## II. COMPUTATIONAL METHODOLOGY

### A. Molecular dynamics with quantum thermal bath

The QTB incorporates the NQEs via the Langevin equation with a time-correlated random force. The Langevin equation is expressed as

$$m_{i\alpha}\frac{d^2 x_{i\alpha}}{dt^2} = -\frac{\partial U}{\partial x_{i\alpha}} - m_{i\alpha}\gamma\frac{dx_{i\alpha}}{dt} + R_{i\alpha}(t), \qquad (2)$$

where $x_{i\alpha}$ and $m_{i\alpha}$ are the three-dimensional cartesian coordinate ($\alpha$ = 1, 2, 3) and mass of the *i*th atom, respectively. $U$, $\gamma$, and *t* are the potential energy, the friction coefficient, and the time, respectively. $R_{i\alpha}(t)$ is the random force acting on the $\alpha$th coordinate of the *i*th atom at time *t*. The second and third terms of the right-hand side in the equation control the temperature, which are fictious forces corresponding to the friction and random forces, respectively. In the standard Langevin thermostat, i.e., the classical thermal bath (CTB), a time-uncorrelated random force is generally used, i.e., *white noise*, whose power spectral density (PSD) is given by

$$I_{R_{i\alpha}}(\omega) = 2m_{i\alpha}\gamma k_B T, \tag{3}$$

where $\omega$ is the angular frequency. The PSD of a random force is independent of the angular frequency, and linearly dependent on the temperature. Using the CTB, the system evolves according to the equipartition theorem, where the mean kinetic energy for each degree of freedom is $(1/2)k_B T$. By contrast, the PSD of a random force in the QTB has frequency dependence, given by

$$I_{R_{i\alpha}}(\omega) = 2m_{i\alpha}\gamma \left( \frac{\hbar|\omega|}{2} + \frac{\hbar|\omega|}{\exp\left(\frac{\hbar|\omega|}{k_B T}\right) - 1} \right), \tag{4}$$

where $\hbar$ is the Dirac constant. This random force called *colored noise* reproduces the mean kinetic and potential energies of lattice vibration according to quantum statistics, which is based on the harmonic approximation.

## B. Jump frequency

According to the TST, the atomic jump frequency $\Gamma$ in an $N$-atom system is given by

$$\Gamma = \sqrt{\frac{k_B T}{2\pi m}} \frac{\int_S \exp\left(-\frac{U}{k_B T}\right) dx^{(3N-1)}}{\int_V \exp\left(-\frac{U}{k_B T}\right) dx^{(3N)}}, \tag{5}$$

where $m$ is mass of diffusion species and $x^{(3N)}$ is the 3$N$-dimenstional coordinate in the configuration space. The numerator is the integration in the range of the 3$N$-dimentional hyper-volume, V, corresponding to the initial state of the atom jump, while the integration range of the denominator, S, is the (3$N$–1)-dimensional saddle hyper-surface separating the initial and final states.

Based on the TST within the classical harmonic approximation [20], the proton jump frequency $\Gamma_{cl}^{h\text{-TST}}$ is approximated by

$$\Gamma_{cl}^{h\text{-TST}} = \frac{\prod_{i=1}^{3N} \nu_i^{\text{initial}}}{\prod_{i=2}^{3N} \nu_i^{\text{saddle}}} \exp\left(-\frac{\Delta U}{k_B T}\right), \tag{6}$$

where $\nu_i^{\text{initial}}$ and $\nu_i^{\text{saddle}}$ are the $i$th eigenfrequencies of the lattice vibrations in the initial and the saddle-point states, respectively. Note that the imaginary eigenfrequency ($i = 1$) at the saddle point state is excluded from the numerator in the pre-exponential factor. Eq. (6) can be rewritten using the classical free energy of lattice vibrations in the initial and saddle-point states, $F_{\text{vib,cl}}^{\text{initial}}$ and $F_{\text{vib,cl}}^{\text{saddle}}$, as follows:

$$\Gamma_{\text{cl}}^{\text{h-TST}} = \frac{k_{\text{B}}T}{h} \exp\left(-\frac{\Delta F_{\text{vib,cl}}^{\text{s-i}}}{k_{\text{B}}T}\right) \exp\left(-\frac{\Delta U}{k_{\text{B}}T}\right), \qquad (7)$$

$$\Delta F_{\text{vib,cl}}^{\text{s-i}} = F_{\text{vib,cl}}^{\text{saddle}} - F_{\text{vib,cl}}^{\text{initial}}, \qquad (8)$$

$$F_{\text{vib,cl}}^{\text{initial}} = \sum_{i=1}^{3N} k_{\text{B}}T \ln\left(\frac{h\nu_i^{\text{initial}}}{k_{\text{B}}T}\right), \qquad (9)$$

$$F_{\text{vib,cl}}^{\text{saddle}} = \sum_{i=2}^{3N} k_{\text{B}}T \ln\left(\frac{h\nu_i^{\text{saddle}}}{k_{\text{B}}T}\right), \qquad (10)$$

where $\Delta F_{\text{vib,cl}}^{\text{s-i}}$ is defined as the change in vibrational free energy from the initial to the saddle-point state. Considering the lattice vibrations within the quantum harmonic approximation [36], the jump frequency is expressed as the similar equation to Eq. (7) using the change in vibrational free energy based on quantum statistics from the initial to the saddle-point state, $\Delta F_{\text{vib,q}}^{\text{s-i}} = F_{\text{vib,q}}^{\text{saddle}} - F_{\text{vib,q}}^{\text{initial}}$,

$$\Gamma_{\text{q}}^{\text{h-TST}} = \frac{k_{\text{B}}T}{h} \exp\left(-\frac{\Delta F_{\text{vib,q}}^{\text{s-i}}}{k_{\text{B}}T}\right) \exp\left(-\frac{\Delta U}{k_{\text{B}}T}\right). \qquad (11)$$

$F_{\text{vib,q}}^{\text{initial}}$ and $F_{\text{vib,q}}^{\text{saddle}}$ are given by

$$F_{\text{vib,q}}^{\text{initial}} = \sum_{i=1}^{3N} \left[\frac{h\nu_i^{\text{initial}}}{2} + k_{\text{B}}T \ln\left(1 - \exp\left(-\frac{h\nu_i^{\text{initial}}}{k_{\text{B}}T}\right)\right)\right], \qquad (12)$$

$$F_{\text{vib,q}}^{\text{saddle}} = \sum_{i=2}^{3N} \left[\frac{h\nu_i^{\text{saddle}}}{2} + k_{\text{B}}T \ln\left(1 - \exp\left(-\frac{h\nu_i^{\text{saddle}}}{k_{\text{B}}T}\right)\right)\right]. \qquad (13)$$

The first term is the zero-point energy, which is the major origin of the difference in vibrational free energy between quantum and classical statistics. The difference gradually decreases as increasing the temperature, and a quantum vibrational free energy coincides with a classical one at the high-temperature limit.

## C. Diffusion coefficient

Once all the frequencies of possible atomic jumps in a given system are evaluated, the diffusion coefficient tensor can be estimated by numerically solving the master equation. The master equation is given by

$$\frac{dp_i(t)}{dt} = \sum_j [\Gamma_{ji} p_j(t) - \Gamma_{ij} p_i(t)], \quad (14)$$

where $p_i(t)$ is the probability of the diffusion species existing at site $i$ as a function of time $t$, and $\Gamma_{ij}$ is the jump frequency from site $i$ to site $j$. The first and second terms of the right-hand side correspond to the inflow and outflow at site $i$, respectively. This equation can readily be solved in the Fourier space [37]. Although the details are omitted here, the components of the diffusion coefficient tensor $D_{mn}$ is obtained as the maximum eigenvalue $\lambda_{\max}$ of the jump matrix $\Lambda$ as follows:

$$\lambda_{\max} = - \sum_{n,m=x,y,z} D_{mn} Q_m Q_n, \quad (15)$$

where $Q_n$ ($n \in \{x,y,z\}$) is the component of the wavenumber vector $\boldsymbol{Q}$. The jump matrix $\Lambda$ is defined by

$$\Lambda_{ij} = \sum_\alpha \left[ \Gamma_{ij}^\alpha \exp(i\boldsymbol{Q} \cdot \boldsymbol{s}_{ij}^\alpha) - \delta_{ij} \sum_{j'} \Gamma_{ij'}^\alpha \right], \quad (16)$$

where $\Gamma_{ij}^\alpha$ and $\boldsymbol{s}_{ij}^\alpha$ are the jump frequency and jump vector from site $i$ to site $j$ in unit cell $\alpha$, and $\delta_{ij}$ is the Kronecker delta. The six independent components in the diffusion coefficient tensor can be estimated by solving the eigenvalue problems of six jump matrices with independent wave number vectors.

## D. Computational conditions

All calculations in the present study were based on the density functional theory (DFT) calculations by the projector augmented wave (PAW) method [38,39] implemented in the Vienna Ab initio Simulation Package (VASP) code [40–42]. The generalized gradient approximation parameterized by Perdew, Burke, Ernzerhof (PBE GGA) was used for the exchange-correlation terms [43]. In the PAW potentials, the 5$s$, 5$p$, 6$s$, and 5$d$ orbitals for Ba, 4$s$, 4$p$, 5$s$, and 4$d$ for Zr, 2$s$ and 2$p$ for O, and 1$s$ for H were treated as valence states. Based on the careful convergence tests, the cut-off energy was first set to 500 eV for optimizing the lattice parameter of the BaZrO$_3$ perfect crystal and was then lowered to 400 eV for all subsequent calculations. The supercell consisting of 2×2×2 BaZrO$_3$ unit cells with a single proton was used as the simulation cell, and the Brillouin zone was sampled with a 2×2×2 grid. An electron was removed from the simulation cell and the uniform background was introduced as counter charge, to reproduce the positive charge of the proton.

MD simulations were performed using on-the-fly machine learning force fields (MLFFs) [33–35] implemented in the VASP code, where the MLFFs are described based on the Smooth Overlap of Atomic Positions (SOAP) [44]. The MLFFs enable long-time MD simulations for sufficient sampling of rare events such as atomic jumps in solids at low temperatures. In the on-the-fly training, we used the NVT ensemble at the constant temperature controlled by the CTB. The time step was set to 1 fs ($1 \times 10^{-15}$ s). Two types of MLFFs were constructed in the present study: one was trained at 1000 K for $5 \times 10^4$ steps and subsequently at 1200 K for $5 \times 10^4$ steps. The other was trained at 2000 K for $3 \times 10^4$ steps and subsequently at 2400 K for $3 \times 10^4$ steps. Figure S1 in the supplementary materials shows a comparison of proton jump frequencies estimated using the constructed MLFFs and the ab-initio MD simulations. The results indicate that the constructed MLFFs give reasonable estimates of proton jump frequencies for this system. In the long-time MD simulations by the MLFFs, The NVT ensemble was employed at the constant temperature in the range of 500 and 2000 K controlled by the QTB or CTB. The former MLFF trained in 1000 and 1200 K was used for the simulations from 500 to 1000 K, and the latter MLFF trained in 2000 and 2400 K was used for the higher temperature simulations. Since the QTB was not implemented in the VASP code, we additionally implemented it in the code [32]. The time step was set to 1 fs ($1 \times 10^{-15}$ s), and the total simulation time was 2 ns ($2 \times 10^6$ steps) after the thermal equilibrium process ($1 \times 10^4$ steps). In both CTB and QTB, the frictional coefficient $\gamma$ was set to 5 THz. In the QTB, a finite support $[-\Omega_{max}, \Omega_{max}]$ of the angular frequency should be determined for the PSD to prevent the divergence of the mean energy. Brrat et al. pointed out that $\Omega_{max}$ should be a few times larger than the highest eigen angular frequency in the system [31]. Therefore, $\Omega_{max}/2\pi$ was set to $2.5 \times 10^{14}$ /s in the present study, which is more than two times larger than the eigenfrequency of the OH stretching mode, $\sim 1 \times 10^{14}$ /s.

Precise structural optimizations were performed for the initial and saddle-point states of each proton jump in BaZrO$_3$. The atomic positions were fully optimized until all atomic force components became smaller than $1 \times 10^{-5}$ [eV/Å]. The potential barriers $\Delta U$ for proton jumps were obtained as the difference in the total energy between the initial and saddle-point states. For estimating the eigenfrequencies, $\nu_i^{initial}$ and $\nu_i^{saddle}$, phonon calculations were performed by the finite-displacement method implemented in the phonopy code [45,46]. The atomic displacements were set to 0.01 Å. The MASTEQ code was used for estimating the proton diffusivity from the estimated jump frequencies based on the master equation [17].

## III. RESULTS AND DISCUSSION

### A. Proton jump frequencies within h-TST

In this subsection, the estimated proton jump frequencies within the h-TST are shown before demonstrating the results of long-time MD simulations. According to the literature [2,14,15], protons in oxides reside around oxide ions and migrate by two kinds of migration paths, i.e., the rotation and hopping paths. Figure 1 shows the reported proton sites and the rotation and hopping paths in the BaZrO$_3$ perfect crystal reported in our previous studies [16,47,48], which was generated by the VESTA software [49]. There are four stable sites around each oxide ion, which are crystallographically equivalent in the perfect crystal. The rotation paths connect these stable sites around single oxide ions, while the hopping paths connect those of the adjacent oxide ions. The calculated potential energy barriers $\Delta U$ of the rotation and hopping paths are 0.16 and 0.26 eV, respectively, which are in good agreement with the reported values in the literature [47]. The calculated potential energy barriers suggest that the hopping path is the rate-limiting step in the proton diffusion without considering the difference in vibrational prefactor, which is consistent with the conventional picture of proton migration in $BaZrO_3$ [2].

Figure 2 shows the proton jump frequencies of the rotation path (black lines) and the hopping path (red lines) estimated from the calculated eigenfrequencies shown in Fig. 3 (a). The dashed and solid lines are the ones estimated based on classical and quantum statistics using Eqs. (7) and (11), respectively. Within the classical h-TST, the vibrational prefactors $\Gamma_0$ of the rotation and hopping paths are 7.0 THz and 35 THz, respectively. Due to the higher $\Gamma_0$ and $\Delta U$ of the hopping path, the two lines of the rotation and the hopping path intersect around 700 K. As a result, the rotation path is the rate-limiting step above 700 K, which is different from the conventional picture. Based on the quantum h-TST, the obtained results become more discrepant from the conventional picture. The jump frequency of the hopping path becomes much higher than the classical one, e.g. ~ 5 times higher at 500 K, while that of the rotation path is almost comparable. The apparent activation energy for the hopping path is 0.18 eV in the range of 500 K and 2000 K, which is almost equal to that for the rotational one, 0.17 eV. This means that proton rotation is the rate-limiting path in the temperature range of interest.

The large difference in the jump frequency of proton hopping between classical and quantum statistics results from the NQEs, particularly zero-point vibration at low temperatures. Focusing on the highest eigenfrequency at each of the initial and the two saddle-point states in Fig. 3(a), both initial state and the saddle-point state of the rotation path have an OH stretching mode with high frequency, 104 THz and 112 THz, respectively. This means that a proton rotates around a single oxide ion without breaking the strong OH bond. By contrast, the highest eigenfrequency is 50.6 THz in the saddle-point state of the hopping path without the OH stretching mode. This is because the proton in the saddle point state of the hopping path is just located on the perpendicular bisector of two adjacent oxide ions, and the direction of proton migration almost coincides with the OH stretching direction. The OH stretching mode therefore has an imaginary frequency in the saddle-point state.

For quantitatively evaluating the contributions of these high-frequency modes to the differences in the proton jump frequencies within the h-TST, Eqs. (7) and (11) are transformed into

$$\frac{\Gamma_q^{\text{h-TST}}}{\Gamma_{\text{cl}}^{\text{h-TST}}} = \exp\left(-\frac{\Delta F_{q-\text{cl}}^{s-i}}{k_B T}\right), \tag{17}$$

where

$$\Delta F_{q-\text{cl}}^{s-i} = \Delta F_{\text{vib},q}^{s-i} - \Delta F_{\text{vib,cl}}^{s-i} = \Delta F_{q-\text{cl}}^{\text{saddle}} - \Delta F_{q-\text{cl}}^{\text{initial}}, \tag{18}$$

$$\Delta F_{q-\text{cl}}^{\text{initial}} = F_{\text{vib},q}^{\text{initial}} - F_{\text{vib,cl}}^{\text{initial}}, \tag{19}$$

$$\Delta F_{q-\text{cl}}^{\text{saddle}} = F_{\text{vib},q}^{\text{saddle}} - F_{\text{vib,cl}}^{\text{saddle}}. \tag{20}$$

$\Delta F_{q-\text{cl}}^{\text{initial}}$ and $\Delta F_{q-\text{cl}}^{\text{saddle}}$ mean the changes in vibrational free energy from classical statistics to quantum statistics at the initial and the saddle-point states, respectively. Figures 4(a) and 4(b) show the contributions of the first, second, third highest, and the other eigenfrequencies to $\Delta F_{q-\text{cl}}^{\text{initial}}$ and $\Delta F_{q-\text{cl}}^{\text{saddle}}$ of the rotation and hopping paths at 500 and 1000 K, where $\Delta F_{q-\text{cl}}^{s-i}$ corresponds to the difference between $\Delta F_{q-\text{cl}}^{\text{saddle}}$ and $\Delta F_{q-\text{cl}}^{\text{initial}}$. In the rotation path, $\Delta F_{q-\text{cl}}^{s-i}$ is small due to the high frequency modes at the saddle-point state of the rotation path comparable to those at the initial state. As a result, the NQEs in the rotation path are almost negligible within the harmonic approximation. In contrast, $\Delta F_{q-\text{cl}}^{s-i}$ is relatively large in the hopping path due to the absence of the highest frequency mode corresponding to the OH stretching mode, resulting in the large NQEs on proton hopping. $\Delta F_{q-\text{cl}}^{s-i}$ at 1000 K are relatively small to those at 500 K for both rotation and hopping paths, which gradually converge to zero with increasing the temperature, leading to $\Gamma_q^{\text{h-TST}} \sim \Gamma_{\text{cl}}^{\text{h-TST}}$.

### B. Proton jump frequencies by MLFF-MD

The proton jump frequencies were estimated from the proton trajectories in the MLFF-MD simulations in two different manners. In the first manner, the proton jump frequency per elementary path, $\Gamma^{\text{count}}$, was estimated by simply counting the number of proton jumps $N_{\text{jump}}$ as follows:

$$\Gamma^{\text{count}} = \frac{N_{\text{jump}}}{n_{\text{path}} t_{\text{tot}}}, \tag{21}$$

where $n_\text{path}$ is the number of equivalent elementary paths per proton site, and $t_\text{tot}$ is the whole simulation time. $\Gamma^\text{count}$ was thus estimated without the two assumptions of the TST and the harmonic approximation. In the second manner, the jump frequency within the TST, $\Gamma^\text{TST}$, was estimated beyond the harmonic approximation in order to verify these two assumptions separately. Setting the proton coordinates as collective variables, $\Gamma^\text{TST}$ was estimated from the existence probability of the proton on the vicinity of the saddle surface S separating the initial and final states as follows:

$$\Gamma^\text{TST} = \frac{N\left(\text{S}\,;\sqrt{\frac{k_\text{B}T}{2\pi m_\text{H}}}\Delta t\right)}{n_\text{path} t_\text{tot}}, \tag{22}$$

where $m_\text{H}$ is the proton mass, and $\Delta t$ is the time step in the MD simulation. $N(\text{S}\,;\Delta x)$ is the number of MD steps at which the distance between the proton and the saddle surface S is less than $\Delta x$. The proton coordinate at any time step was transferred to the single asymmetric unit by symmetry operations. The saddle surfaces S of the rotation and hopping paths correspond to the blue and yellow surfaces in Fig. 1, respectively. Note that $n_\text{path}$ is 2 in the denominators of Eqs. (21) and (22) because there are two crystallographically equivalent paths from a stable site to another with respect to both rotation and hopping paths. Hereafter, $\Gamma^\text{count}$ and $\Gamma^\text{TST}$ estimated from the MD simulations using the CTB and the QTB (CTB-MD and QTB-MD, hereafter) are denoted by $\Gamma_\text{CTB}^\text{count}$, $\Gamma_\text{QTB}^\text{count}$, $\Gamma_\text{CTB}^\text{TST}$ and $\Gamma_\text{QTB}^\text{TST}$, respectively.

Figure 5 shows the proton jump frequencies estimated by the CTB-MD as a function of inverse temperature, in which $\Gamma_\text{cl}^\text{h-TST}$ is also shown by the dashed lines for comparison. $\Gamma_\text{CTB}^\text{count}$, $\Gamma_\text{CTB}^\text{TST}$, and $\Gamma_\text{cl}^\text{h-TST}$ are comparable in the rotation path, where the apparent activation energies are the same, 0.16 eV, and the vibrational prefactors are about 10 THz, i.e., 8.6, 12, and 7.0 THz, respectively. This suggests that the TST and the harmonic approximation are reasonable for the rotation path within the classical approximation. By contrast, $\Gamma_\text{CTB}^\text{count}$, $\Gamma_\text{CTB}^\text{TST}$, and $\Gamma_\text{cl}^\text{h-TST}$ are different in the hopping path, particularly much lower $\Gamma_\text{CTB}^\text{count}$ than $\Gamma_\text{CTB}^\text{TST}$ and $\Gamma_\text{cl}^\text{h-TST}$. The apparent activation energies of $\Gamma_\text{CTB}^\text{count}$, $\Gamma_\text{CTB}^\text{TST}$, and $\Gamma_\text{cl}^\text{h-TST}$ are 0.22, 0.21, and 0.26 eV, respectively, and the vibrational prefactors are 2.9, 12, and 35 THz, respectively. This indicates that the assumptions in the TST overestimate the vibrational prefactor, and that the harmonic approximation overestimates both apparent activation energy and vibrational prefactor. The overestimation by the TST is reasonable because the TST neglects the diffusion trajectories returning to the initial state after crossing the saddle surface S [50]. Without the TST assumptions and the harmonic approximation, the proton hopping is the rate-limiting process in the temperature range of interest within the classical framework. In our previous study, the jump frequencies in the proton rotation and hopping paths were estimated from the free energy surfaces constructed by metadynamics simulations within the TST [51]. The apparent activation energy and the vibrational prefactor are 0.18 eV and 13 THz for the proton rotation and 0.23 eV and 14 THz for the proton hopping, respectively, which are in excellent agreement with $\Gamma_{\text{CTB}}^{\text{TST}}$ estimated from Eq. (21) in the present study.

Figures 6(a) and (b) show the jump frequencies estimated from the QTB-MD for the proton rotation and hopping, respectively, in which $\Gamma_{\text{q}}^{\text{h-TST}}$ is also shown by the solid line for comparison. In the rotation path, $\Gamma_{\text{q}}^{\text{h-TST}}$, $\Gamma_{\text{QTB}}^{\text{TST}}$ and $\Gamma_{\text{QTB}}^{\text{count}}$ are comparable in the high temperature range as in the classical case. However, $\Gamma_{\text{QTB}}^{\text{TST}}$ and $\Gamma_{\text{QTB}}^{\text{count}}$ become larger than $\Gamma_{\text{q}}^{\text{h-TST}}$ as the temperature decreases. In the hopping path, $\Gamma_{\text{QTB}}^{\text{TST}}$ and $\Gamma_{\text{QTB}}^{\text{count}}$ show a strong non-linear behavior in the Arrhenius plot in contrast to $\Gamma_{\text{q}}^{\text{h-TST}}$, where the slope becomes moderate gradually with decreasing temperature. $\Gamma_{\text{QTB}}^{\text{TST}}$ is larger than $\Gamma_{\text{QTB}}^{\text{count}}$ in the whole temperature range of interest, which derives from the same origin as in the difference between $\Gamma_{\text{CTB}}^{\text{TST}}$ and $\Gamma_{\text{CTB}}^{\text{count}}$. Figure 6(c) shows the comparison between $\Gamma_{\text{QTB}}^{\text{count}}$ and $\Gamma_{\text{CTB}}^{\text{count}}$ for the proton rotation and hopping paths. $\Gamma_{\text{QTB}}^{\text{count}}$ is comparable to $\Gamma_{\text{CTB}}^{\text{count}}$ in each path at the highest temperature, 2000 K, but gradually deviates from $\Gamma_{\text{CTB}}^{\text{count}}$ with decreasing the temperature. The hopping path exhibits larger deviation between $\Gamma_{\text{QTB}}^{\text{count}}$ and $\Gamma_{\text{CTB}}^{\text{count}}$ than the rotation path, which can also be seen in the comparison between the quantum and classical h-TST (Fig. 2). As a result, the difference in the jump frequency between the two paths is much smaller in the QTB-MD than that in the CTB-MD in the low temperature range, though the hopping path is the rate-limiting in both the QTB-MD and CTB-MD.

The QTB-MD incorporates the NQEs through the kinetic energies, in which the kinetic energy given to each vibrational mode depends on the eigenfrequency. Particularly, the apparent temperature of the proton, $\tilde{T}_\text{H}$, defined by Eq. (23) should be largely different from the preset temperature, $T^\text{preset}$, because the high-frequency vibrational modes with large zero-point vibration energies are attributed to the extremely light proton.

$$\tilde{T}_\text{H} = \frac{\frac{1}{2} m_\text{H} \overline{v_\text{H}^2}}{\frac{3}{2} k_\text{B}}, \quad (23)$$

where $\overline{v_\text{H}^2}$ is the mean square velocity of the proton. Figure 7 shows the $\tilde{T}_\text{H}$ in the QTB-MD, which is higher than the $T^\text{preset}$ particularly at low temperatures, in contrast to the $\tilde{T}_\text{H}$ in the CTB-MD comparable to the $T^\text{preset}$. This means that the proton is locally heated to the higher temperatures by the QTB. As a result, the probability density of the proton in the QTB-MD ($\rho_\text{QTB}$) has relatively wide distribution to that in the CTB-MD ($\rho_\text{CTB}$). Figure 8 shows the $\rho_\text{CTB}$ and the $\rho_\text{QTB}$ at 500 and 1000 K on the cross-sectional planes including the rotation and hopping paths. Their difference, i.e., $\rho_\text{QTB} - \rho_\text{CTB}$, is also shown in the figure for clear comparison. Overall, $\rho_\text{QTB} - \rho_\text{CTB}$ is negative around the proton site and positive in the vicinity of the saddle surface S, leading to the larger $\Gamma_\text{QTB}^\text{count}$ than $\Gamma_\text{CTB}^\text{count}$. The difference between $\rho_\text{CTB}$ and $\rho_\text{QTB}$ is more pronounced at lower temperatures, and more significant in the hopping path than in the rotation path, resulting in the larger $\Gamma_\text{QTB}^\text{count}$ than $\Gamma_\text{CTB}^\text{count}$ at low temperatures in the hopping path.

Figure 9 compares the estimated proton jump frequencies in the present study with those in two previous studies [25–28]. The first one was reported by Zhang et al. [25], where the proton jump frequencies were estimated by path-integral molecular dynamics within the path-centroid transition state theory [52]. In their study, the proton jump frequencies without NQEs were also estimated within the classical TST. In Fig. 9, their estimated jump frequencies with and without NQEs are denoted by the solid and open green symbols, respectively. Neglecting the NQEs, the apparent activation energies in the rotation and hopping paths are reported to be 0.17 and 0.22 eV, respectively, which are comparable to those of $\Gamma_\text{CTB}^\text{TST}$ in the present study, i.e., 0.16 and 0.21 eV, respectively. Their exponential prefactors in the rotation and hopping paths are 16 and 40 THz, respectively, while those of $\Gamma_{\text{CTB}}^{\text{TST}}$ in the present study are both 12 THz. The difference in the exponential prefactor in the hopping path is relatively large, which is probably due to the difference in the simulation cell size, a $1 \times 1 \times 1$ unit cell in their study vs. a $2 \times 2 \times 2$ supercell in the present study. Considering the NQEs, their estimated jump frequencies are different from those without NQEs even at 600 K, and the difference increases with decreasing temperature, particularly in proton hopping. The two high-temperature extrapolations of their hopping frequencies with and without the NQEs intersect around 850 K, suggesting that their hopping frequency should exhibit non-linear behavior with slope change around the intersection. The non-linear behavior is consistent with $\Gamma_{\text{QTB}}^{\text{TST}}$ in the present study, where the slope change is observed around 1000 K. The apparent activation energies below 600 K are 0.07 eV in the present study and 0.08 eV in their study, both of which are in excellent agreement. Although the QTB-MD neglects the tunneling effects, this coincidence is not surprising because Zhang et al. claimed that the proton rotation and hopping can be considered as the semiclassical over barrier motion in the temperature range from 300 to 600 K [25].

The similar trend also suggests that a well-known issue in QTB-MD, i.e., the zero-point energy leakage (ZPEL) [30,53,54], is negligible for evaluating the proton jump frequencies in the intermediate temperature range. The ZPEL means that the zero-point energies of high-frequency vibrational modes are partially transferred to low-frequency modes in QTB-MD, leading to a wrong energy distribution in terms of statistical mechanics. In this system, the OH stretching and bending modes have relatively high eigenfrequencies (see Fig. 3(a)), so that the zero-point energies are partially transferred to the host lattice vibration with lower eigenfrequencies. As a result, the host lattice is overheated during the QTB-MD, which may accelerate the proton jump. Figure S2 shows the classical temperatures of the host lattice and the single proton converted from the mean kinetic energies in the QTB-MD. The classical temperatures of the host lattice and the proton are higher and lower than expected from the preset temperature under the quantum harmonic approximation, respectively. However, the temperature increase in the host lattice is 40 K at most, and the non-linear behavior in the hopping frequency in the QTB-MD remains observed even in the Arrhenius plots using the host lattice effective temperature (the blue solid triangles in Fig. S3). Taking the cooled proton by the ZPEL into consideration, the blue solid triangles correspond to the lower limit by the ZPEL correction. The upward deviation of $\Gamma_{\text{QTB}}^{\text{TST}}$ from $\Gamma_{\text{CTB}}^{\text{TST}}$ at low temperatures is therefore significant in proton hopping, although the deviation is slight in proton rotation.

The blue lines in Fig. 9 show the estimated jump frequencies by Geneste, where Ref. [28] and Ref. [27] were referred for the proton rotation and hopping, respectively. In his study, the proton jump frequencies were estimated in a wider temperature range from 150 to 900 K within the TST. The average flux crossing the saddle hypersurface was estimated from the probability of the adiabatic and the non-adiabatic proton tunneling at coincidence configurations at which the energy levels in the initial and final states coincide with each other. In proton rotation, his jump frequencies are in reasonable agreement with those reported by Zhang et al. and $\Gamma_{\text{QTB}}^{\text{TST}}$ in the present study. By contrast, his jump frequencies of proton hopping are smaller than those reported by Zhang et al. and $\Gamma_{\text{QTB}}^{\text{TST}}$ in the present study. The apparent activation energy is 0.13 eV, larger than those in the present study and in Zhang's study. One of possible reasons for this difference is that a proton was explicitly treated as a quantum particle having discrete energy levels in his research. This rigorous quantum nature of the proton is described by assuming that the proton quantum state immediately follows the motion of the host lattice, i.e., the Born-Oppenheimer approximation. In addition, the structural change in the host crystal lattice was restricted only to two degrees of freedom, i.e., *reorganization* and *facilitating step*, and the diabatic surfaces in the two-dimensional configuration space was expanded quadratically. The framework of this two-lattice-vibration model in Geneste's study is totally different from our study and Zhang's study, possibly leading to the different trend in the Arrhenius plots of the proton jump frequencies.

## C. Proton diffusivity and isotope effects

The jump frequencies of the proton rotation and hopping were converted into the proton diffusion coefficients based on the master equation [17]. Figure 10 shows the estimated proton diffusion coefficients by (a) the CTB-MD and (b) the QTB-MD, in which the diffusion coefficients estimated from the mean square displacements, $D^{\text{MSD}}$, are also shown for verifying Markovianity in the master equation.

$$D^{\text{MSD}} = \lim_{t \to \infty} \frac{\langle (x(t) - x(0))^2 \rangle}{6t}, \qquad (24)$$

where $x(t)$ is the position of proton at time $t$, and $\langle X \rangle$ is the expected value of random variable $X$. $D^{\text{MSD}}$ were estimated at 500, 1000, and 2000 K from the mean square displacements of 50 iterations. The simulation times were set to 2, 1, and 1 ns, respectively, which are enough long for protons to migrate over a long range, e.g., the root mean square displacement is 24 Å at 500 K in the CTB-MD. The super- and sub-scripts of diffusion coefficients have the same meanings as those of jump frequencies, e.g., the diffusion coefficient estimated from the jump frequencies $\Gamma_{\text{cl}}^{\text{h-TST}}$ is denoted by $D_{\text{cl}}^{\text{h-TST}}$. $D^{\text{MSD}}$ in the CTB-MD and the QTB-MD are denoted by $D_{\text{CTB}}^{\text{MSD}}$ and $D_{\text{QTB}}^{\text{MSD}}$, respectively. In the classical treatment without the NQEs, the diffusion coefficients are almost linear in the Arrhenius plots. $D_{\text{cl}}^{\text{h-TST}}$ is comparable to $D_{\text{CTB}}^{\text{TST}}$ reflecting the relatively small anharmonicity. However, $D_{\text{CTB}}^{\text{TST}}$ is more than 3 times larger than $D_{\text{CTB}}^{\text{count}}$ mainly because the jump frequency is generally overestimated by the TST. In the quantum treatment with the NQEs, the diffusion coefficients except $D_{\text{q}}^{\text{h-TST}}$ exhibit non-Arrhenius behaviors. $D_{\text{QTB}}^{\text{TST}}$ deviates from $D_{\text{q}}^{\text{h-TST}}$ with decreasing temperature, directly reflecting that $\Gamma_{\text{QTB}}^{\text{TST}}$ has a relatively moderate slope in the Arrhenius plot to $\Gamma_{\text{q}}^{\text{h-TST}}$ at low temperatures (Fig. 6). $D_{\text{QTB}}^{\text{count}}$ is smaller than $D_{\text{QTB}}^{\text{TST}}$ due to the overestimation by the TST as seen in the classical treatment. $D^{\text{MSD}}$ coincides with $D^{\text{count}}$ within the computational accuracy in both CTB-MD and QTB-MD. This suggests that the proton jumps in BaZrO$_3$ can be regarded as a Markovian process and that the diffusivity estimation based on the master equation is reasonable in this system. Figure 10(c) shows the comparison between $D_{\text{CTB}}^{\text{count}}$ and $D_{\text{QTB}}^{\text{count}}$, which are comparable at high temperatures due to the negligible NQEs. However, $D_{\text{QTB}}^{\text{count}}$ deviates upwards from $D_{\text{CTB}}^{\text{count}}$ as the temperature decreases, e.g. 1.6 and 8.4 times larger at 1000 and 500 K, respectively. This suggests that the proton diffusivity is strongly affected by the NQEs even at intermediate temperatures (573–

873 K).

Finally, the isotope effects on the proton/deuteron diffusivity are discussed, which are widely used as a convenient method to make sure the proton conductivity. The conventional understanding of the isotope effects should be updated by taking the NQEs and the anharmonicity into account. Conventionally, the ratio of the proton and deuteron diffusion coefficients, $r_D = D_H/D_D$, which is equal to that of the proton and deuteron conductivity $\sigma_H/\sigma_D$ at the same concentration, is considered to be approximately the square root of mass ratio, $\sqrt{m_D/m_H}(=\sqrt{2})$, or slightly lower, which is originated from Eq. (6) based on the classical h-TST [20]. However, $\sigma_H/\sigma_D$ is experimentally reported to deviate from $\sqrt{2}$ significantly in many proton conducting oxides [55], which is considered to be mainly attributed to the difference in zero-point energies of proton and deuteron vibrations within the framework of the quantum h-TST.

Figures 11(a) and (b) show the H/D ratios of the estimated jump frequencies in the rotation and hopping paths, respectively. The dashed and solid lines represent the H/D ratios of $\Gamma_{cl}^{h\text{-}TST}$ and $\Gamma_{q}^{h\text{-}TST}$, denoted by $r_{\Gamma,cl}^{h\text{-}TST}$ and $r_{\Gamma,q}^{h\text{-}TST}$ respectively. $\Gamma_{cl}^{h\text{-}TST}$ and $\Gamma_{q}^{h\text{-}TST}$ of deuterons were estimated simply by replacing the proton mass by the deuteron mass in the dynamical matrices for estimating the eigenfrequencies. In the classical treatment, $r_{\Gamma,cl}^{h\text{-}TST}$ is comparable to the root of the reduced mass ratio of O-H and O-D, i.e., $\sqrt{17/9} \sim 1.374$. In the rotation path, $r_{\Gamma,q}^{h\text{-}TST}$ is almost equal to $r_{\Gamma,cl}^{h\text{-}TST}$ and independent of the temperature. This is because the NQEs are almost negligible in the rotation path within the h-TST as discussed in Sec. IIIA. On the other hand, in the hopping path, $r_{\Gamma,q}^{h\text{-}TST}$ is different from $r_{\Gamma,cl}^{h\text{-}TST}$ and increases with decreasing the temperature. In this case, $\Gamma_{q}^{h\text{-}TST}$ is higher than $\Gamma_{cl}^{h\text{-}TST}$ due to the absence of the OH (OD) stretching mode with the highest eigenfrequency in the saddle-point state (Fig. 3). Since the eigenfrequency of the OH stretching mode is higher than that of the OD stretching mode, $\Gamma_{q}^{h\text{-}TST}/\Gamma_{cl}^{h\text{-}TST}$ of the proton jump is higher than that of deuteron jump in comparison at the same temperature (Figs. 3 and 4). This results in relatively large $r_{\Gamma,q}^{h\text{-}TST}$ to $r_{\Gamma,cl}^{h\text{-}TST}$, which is more prominent at lower temperatures.

The open and solid symbols in Figs. 11(a) and (b) denote $r_{\Gamma,\text{CTB}}^{\text{count}}$ and $r_{\Gamma,\text{QTB}}^{\text{count}}$, i.e., the H/D ratios of $\Gamma_{\text{CTB}}^{\text{count}}$ and $\Gamma_{\text{QTB}}^{\text{count}}$, respectively. The MLFFs constructed for proton were employed for the deuteron MD simulations. Note that there are intrinsic errors in $r_{\Gamma,\text{CTB}}^{\text{count}}$ and $r_{\Gamma,\text{QTB}}^{\text{count}}$ derived from the counted numbers of proton and deuteron jumps in the MD simulations, $N_{\text{jump}}$, whose standard deviation is $\sqrt{N_{\text{jump}}}$ corresponding to that of Poisson distribution. The error bars in Fig. 11 were estimated according to the error propagation formula using twice the standard deviation of the counted number of jumps. Since the error is significant at lower temperatures due to the small $N_{\text{jump}}$, the simulation times were set to 6 ns at 500 and 550 K, and 4 ns at 600, 650 and 700 K. $r_{\Gamma,\text{CTB}}^{\text{count}}$ is comparable to $r_{\Gamma,\text{cl}}^{\text{h-TST}}$ in the rotation path, while $r_{\Gamma,\text{CTB}}^{\text{count}}$ is a little smaller than $r_{\Gamma,\text{cl}}^{\text{h-TST}}$ in the hopping path. $r_{\Gamma,\text{QTB}}^{\text{count}}$ increases with decreasing the temperature in both rotation and hopping paths. This trend is more significant in the rotation path, which is different from the trend in $r_{\Gamma,\text{q}}^{\text{h-TST}}$.

Figure 11(c) shows the H/D ratio of diffusion coefficients estimated by the master equation, where the dashed and solid lines denote $r_{D,\text{cl}}^{\text{h-TST}}$ and $r_{D,\text{q}}^{\text{h-TST}}$ obtained from $\Gamma_{\text{cl}}^{\text{h-TST}}$ and $\Gamma_{\text{q}}^{\text{h-TST}}$, respectively. $r_{D,\text{cl}}^{\text{h-TST}}$ is constant around 1.38, which is consistent with $r_{\Gamma,\text{cl}}^{\text{h-TST}}$. On the contrary, $r_{D,\text{q}}^{\text{h-TST}}$ increases with decreasing the temperature, resulting from the temperature dependence of $r_{\Gamma,\text{q}}^{\text{h-TST}}$ in the hopping path. Since the hopping path is not rate-limiting step within the h-TST in the quantum statistics (Fig. 2), $r_{D,\text{q}}^{\text{h-TST}}$ is smaller than $r_{\Gamma,\text{q}}^{\text{h-TST}}$ in the hopping path, e.g., $r_{D,\text{q}}^{\text{h-TST}} = 1.66$ and $r_{\Gamma,\text{q}}^{\text{h-TST}} = 2.48$ at 500 K. In contrast, $r_{D,\text{CTB}}^{\text{count}}$ is almost equal to $r_{\Gamma,\text{CTB}}^{\text{count}}$ in the hopping path, because the hopping path is rate-limiting in $\Gamma_{\text{CTB}}^{\text{count}}$ in this temperature range (Fig. 5). $r_{D,\text{QTB}}^{\text{count}}$ becomes larger as the temperature decreases, which reaches as high as 1.9 at 500 K. This trend is consistent with experimental isotope effects reported by Nowick and Vaysleyb [55], where the H/D ratio of conductivity in many kinds of oxides are from 1.5 to 3 at 500 K. However, it should be noted that the effects of dopants and the difference in concentration between protons and deuterons were neglected in the present study.

## IV. Conclusions

In the present study, the NQEs on proton diffusivity in $BaZrO_3$ were investigated by the MD simulations with the QTB using the on-the-fly machine learning force field. The jump frequencies of proton rotation and hopping were evaluated by the MD simulations with the QTB and the CTB, which were also evaluated within the h-TST by the DFT-based phonon calculations. The validation of the conventional classical h-TST for quantitative evaluation of proton diffusivity in oxides were then examined, where the TST, the harmonic approximation, and the classical approximation were separately verified.

Within the classical h-TST, the proton rotation is rate-limiting above 700 K and the proton hopping is rate-limiting at the lower temperatures. Evaluating the NQEs within the h-TST, the NQEs are significant in the hopping path while almost negligible in the rotation path. The difference results from the absence or presence of the highest frequency mode (OH stretching mode) at the saddle-point state. Within the h-TST, the NQEs makes the proton rotation rate-limiting in the whole temperature range of interest (500–2000 K).

The jump frequencies estimated from the CTB-MD in the two different manners clarified that the h-TST is reasonable for the proton rotation but unreasonable for the proton hopping within the classical approximation. Specifically, the TST overestimates the pre-exponential factor of the hopping frequency, and the harmonic approximation overestimates both the preexponential factor and the activation energy, resulting in the proton hopping is rate-limiting in the whole temperature range of interest (500–2000 K).

The jump frequencies estimated from the QTB-MD show a non-Arrhenius behavior, particularly in the proton hopping, where the slope in the Arrhenius plot becomes moderate with decreasing temperature. As a result, the proton diffusion coefficient has the similar behavior, suggesting the importance of the NQEs in the proton diffusivity in oxides even at the intermediate temperatures (573–873 K). The isotope effect is one of examples showing the importance of the NQEs. In the classical treatment, the H/D ratio of the diffusion coefficient is independent of temperature and equal to or a little smaller than $\sqrt{2}$, reflecting the square root of their mass ratio. However, the ratio becomes larger with decreasing temperature, which reaches 1.9 at 500 K.

**Acknowledgments**

This paper was supported by JST, PRESTO (Grant No. JPMJPR24J8) and JSPS, KAKENHI (Grant No. 24K01147). KK was financially supported by JST, SPRING (Grant No. JPMJSP2110) and Fujinomori-Masamichi Memorial Scholarship in the Mining and Materials Processing Institute of Japan.

**Figure captions**

FIG. 1. The proton stable sites (white spheres) and the rotation and hopping paths (black and white lines, respectively) in the BaZrO$_3$ crystal, which are shown in the supercell consisting of $2 \times 2 \times 2$ unitcells. The green, grey, and red spheres denote Ba, Zr and O ions, respectively. The colored tetrahedron denotes the asymmetric unit, in which the blue and yellow surfaces correspond to the saddle surfaces for proton rotation and hopping, respectively. This figure was generated by the VESTA software [49].

FIG. 2. The estimated proton jump frequencies within the h-TST. The black and red lines denote the rotation and hopping paths, respectively. The dashed and solid lines denote $\Gamma_{\text{cl}}^{\text{h-TST}}$ and $\Gamma_{\text{q}}^{\text{h-TST}}$ based on the classical and quantum statistics, respectively.

FIG. 3. Estimated eigenfrequencies of lattice vibrations at the initial state and the saddle-point states in the rotation and hopping paths for the BaZrO$_3$ crystal with (a) a single proton and (b) a single deuteron. The red, blue and green lines denote the first, second and third highest eigenfrequencies in each state, respectively. The dashed lines denote the imaginary frequencies represented by minus-sign frequencies.

FIG. 4. Contributions of the three highest and other lower eigenfrequencies to the difference in vibrational free energy at the initial state and the saddle-point states of the rotation and hopping paths between the quantum and classical statistics, i.e., $\Delta F_{\text{q-cl}}^{\text{initial}}/k_{\text{B}}T$ and $\Delta F_{\text{q-cl}}^{\text{saddle}}/k_{\text{B}}T$ in Eq. (19) and (20), respectively. (a) and (b) correspond to the case of proton at 500 and 1000 K, respectively, while (c) and (d) are the case of deuteron at 500 and 1000 K, respectively. The red, blue, green and grey accumulated bars denote the first, second, third highest, and other lower eigenfrequencies, respectively, corresponding to the eigenfrequencies shown in Fig. 3.

FIG. 5. Proton jump frequencies estimated by CTB-MD in (a) the rotation path and (b) the hopping path. The open circles denote the counted jump frequencies during the MD simulations. The open triangles denote the jump frequencies estimated within the TST but beyond the harmonic approximation. The dashed lines are the estimated proton jump frequencies under the classical h-TST.

FIG. 6. Proton jump frequencies estimated by the QTB-MD in (a) the rotation path and (b) the hopping path. The solid circles denote the counted jump frequencies during the MD simulations. The solid triangles denote the jump frequencies within the TST but beyond the harmonic approximation. The solid lines are the estimated proton jump frequencies under the quantum h-TST. (c) Comparison in the counted proton jump frequency between the CTB-MD (open symbols) and the QTB-MD (solid symbols).

FIG. 7. The apparent temperature of the proton as a function of preset temperature. The open circles and solid triangles denote the apparent temperature in the CTB-MD and the QTB-MD, respectively. The dotted line denotes the preset temperature.

FIG. 8. The left two columns show the probability densities of the proton at 500 and 1000 K obtained from the CTB-MD and the QTB-MD denoted by $\rho_{\text{CTB}}$ and $\rho_{\text{QTB}}$, respectively. The right column shows the difference in the probability density between the CTB-MD and the QTB-MD, $\rho_{\text{QTB}} - \rho_{\text{CTB}}$. The probability densities are shown on the cross-sectional planes including the rotation and hopping paths. The green, grey and red circles denote Ba, Zr and O ions, respectively. The probability density is normalized in the asymmetric unit (Fig. 1), i.e., $\int_{\text{assym. unit}} \rho \, d\boldsymbol{x} = 1$.

FIG. 9. The proton jump frequencies of (a) proton rotation and (b) proton hopping estimated in the present study and in the previous theoretical studies by Zhang et al. [25] and Geneste [27,28]. The open and solid triangles denote the $\Gamma^{\mathrm{TST}}$ defined by Eq. (22) in the CTB-MD and the QTB-MD. respectively. The green symbols denote those estimated by Zhang et al. using the PIMD, where the open and solid symbols correspond to those without and with the NQEs, respectively. The blue lines denote those estimated by Geneste assuming the adiabatic and non-adiabatic proton tunneling.

FIG. 10. Diffusion coefficients estimated by (a) the CTB-MD and (b) the QTB-MD. The triangles and circles denote the diffusion coefficients estimated within or beyond the TST, respectively. The diffusion coefficients under the h-TST based on the classical and quantum statistics are also shown by the dashed and solid lines, respectively. The blue squares denote those estimated from the mean square displacements. (c) Comparison between the diffusion coefficients estimated by the CTB-MD and the QTB-MD, shown by the open and solid symbols, respectively.

FIG. 11. H/D ratios of the estimated jump frequencies in (a) the rotation path and (b) the hopping path, and (c) those of diffusion coefficients. The open and solid symbols correspond to those estimated from the CTB-MD and the QTB-MD, respectively. The dashed and solid lines denote those within the classical and quantum h-TST, respectively. The error bars were estimated according to the error propagation formula based on twice the standard deviation of the counted jump number of protons or deuterons.

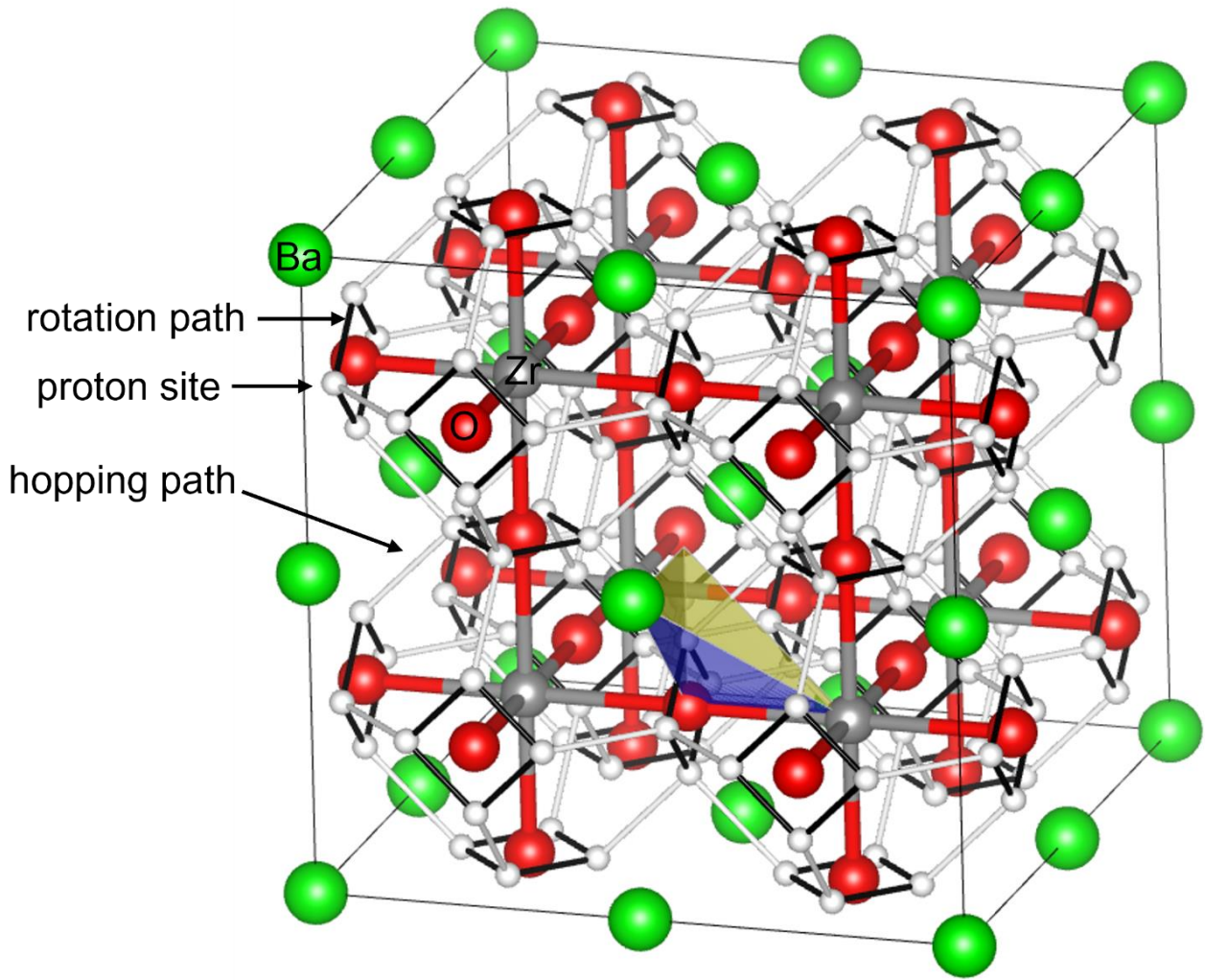

FIG. 1. The proton stable sites (white spheres) and the rotation and hopping paths (black and white lines, respectively) in the BaZrO$_3$ crystal, which are shown in the supercell consisting of $2 \times 2 \times 2$ unitcells. The green, grey, and red spheres denote Ba, Zr and O ions, respectively. The colored tetrahedron denotes the asymmetric unit, in which the blue and yellow surfaces correspond to the saddle surfaces for proton rotation and hopping, respectively. This figure was generated by the VESTA software [49].

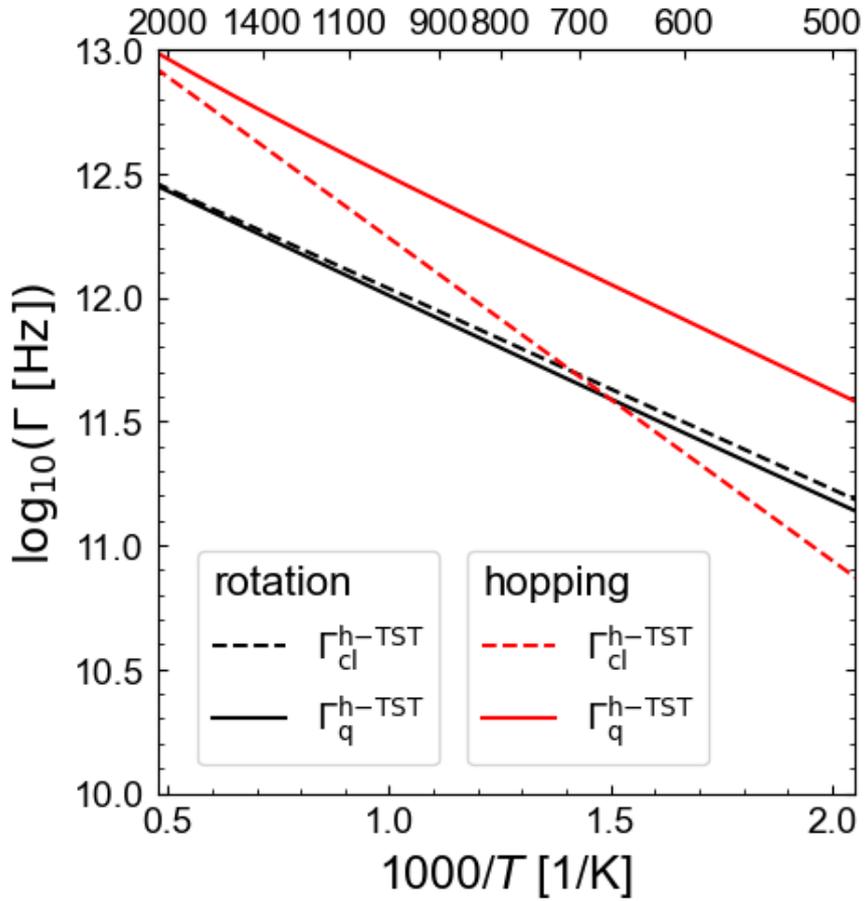

FIG. 2. The estimated proton jump frequencies within the h-TST. The black and red lines denote the rotation and hopping paths, respectively. The dashed and solid lines denote $\Gamma_{\text{cl}}^{\text{h-TST}}$ and $\Gamma_{\text{q}}^{\text{h-TST}}$ based on the classical and quantum statistics, respectively.

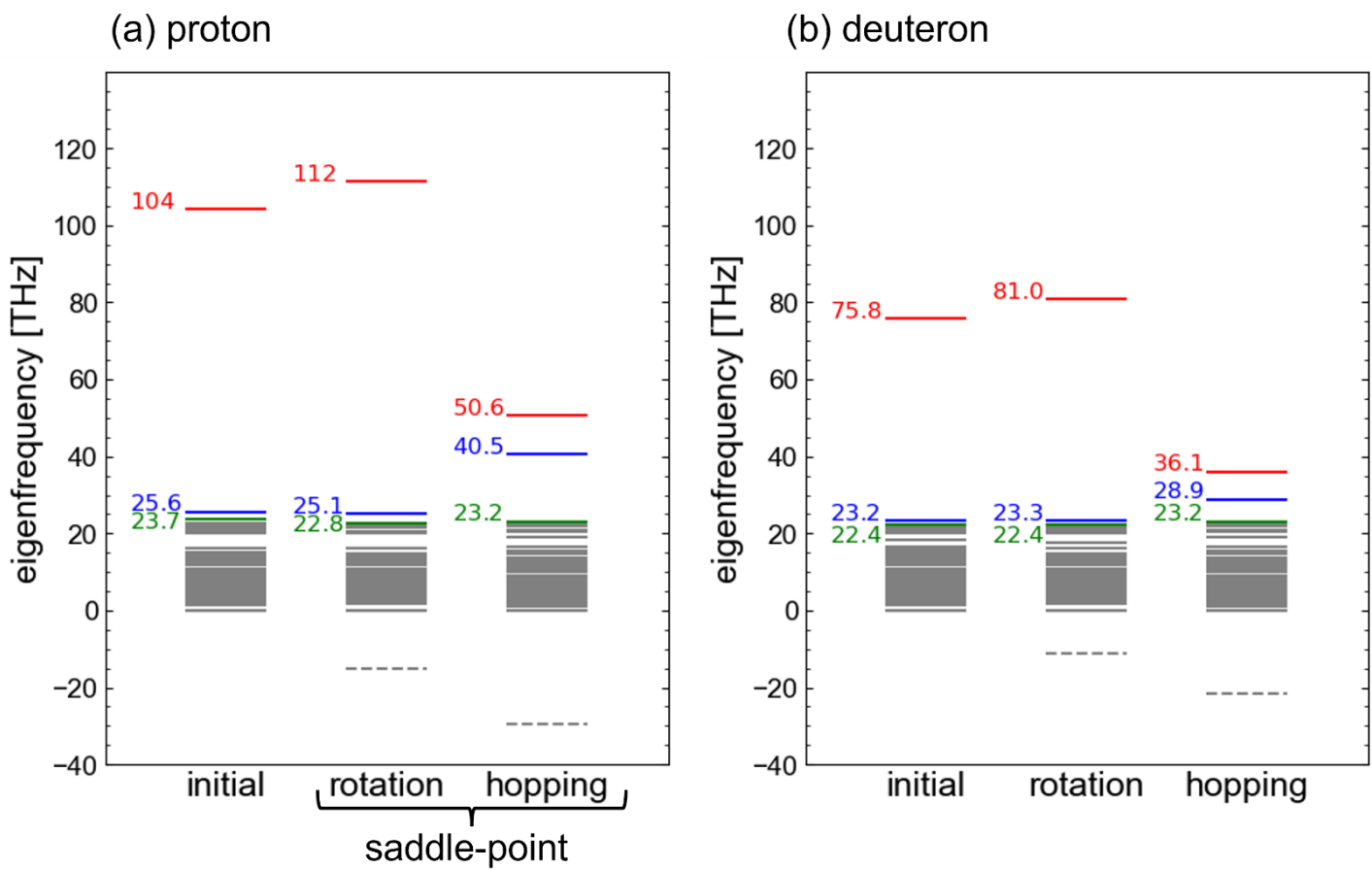

FIG. 3. Estimated eigenfrequencies of lattice vibrations at the initial state and the saddle-point states in the rotation and hopping paths for the $BaZrO_3$ crystal with (a) a single proton and (b) a single deuteron. The red, blue and green lines denote the first, second and third highest eigenfrequencies in each state, respectively. The dashed lines denote the imaginary frequencies represented by minus-sign frequencies.

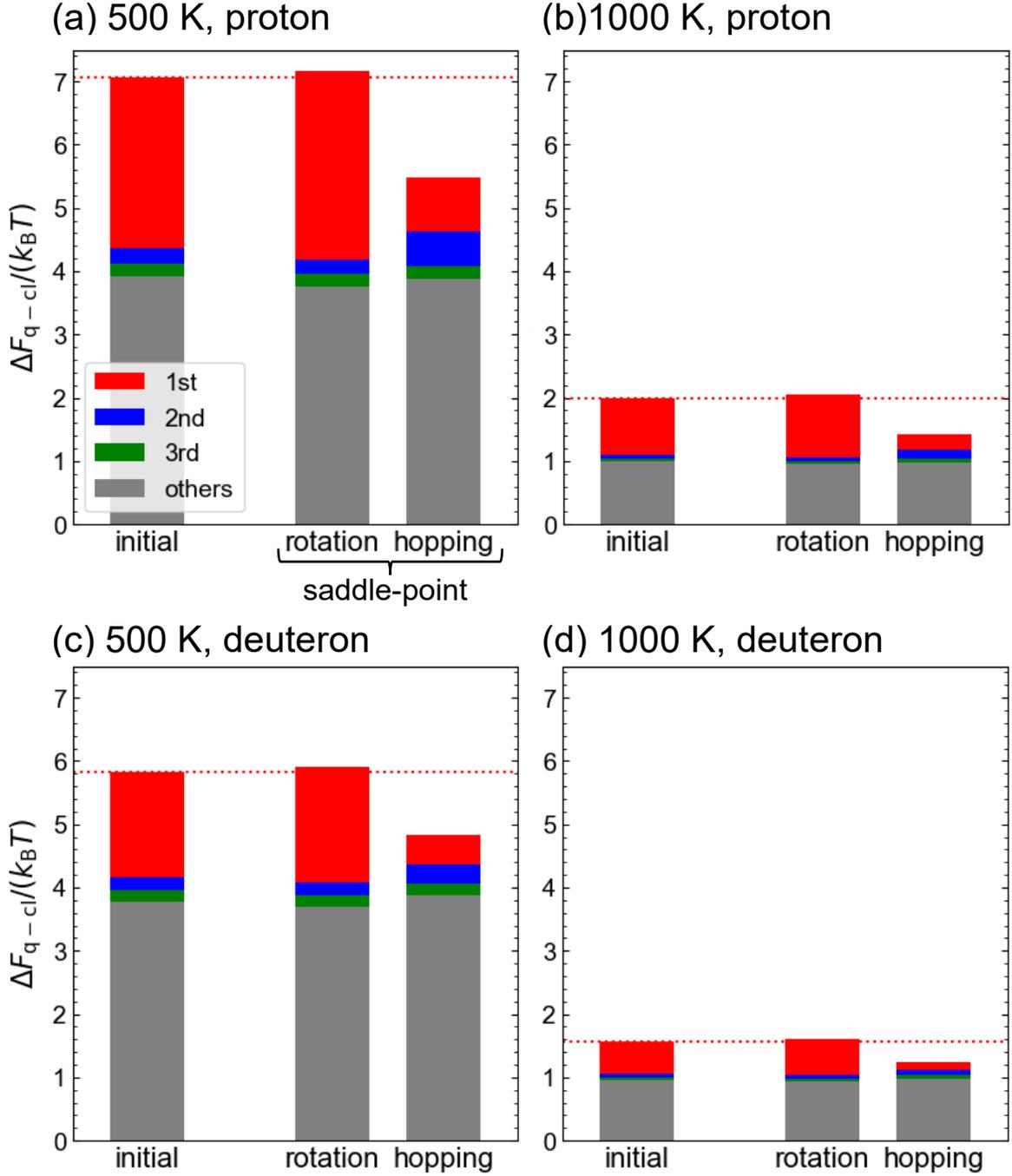

FIG. 4. Contributions of the three highest and other lower eigenfrequencies to the difference in vibrational free energy at the initial state and the saddle-point states of the rotation and hopping paths between the quantum and classical statistics, i.e., $\Delta F_{q-cl}^{initial}/k_B T$ and $\Delta F_{q-cl}^{saddle}/k_B T$ in Eq. (19) and (20), respectively. (a) and (b) correspond to the case of proton at 500 and 1000 K, respectively, while (c) and (d) are the case of deuteron at 500 and 1000 K, respectively. The red, blue, green and grey accumulated bars denote the first, second, third highest, and other lower eigenfrequencies, respectively, corresponding to the eigenfrequencies shown in Fig. 3.

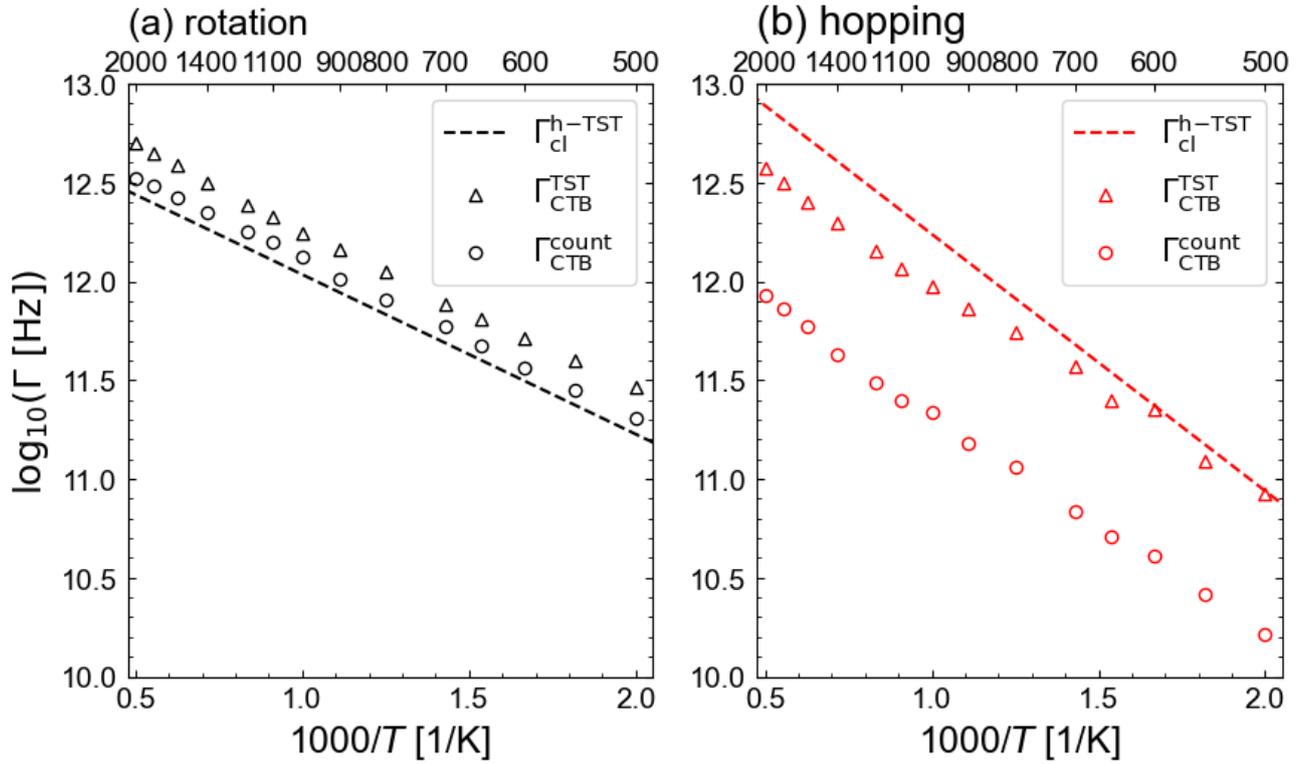

FIG. 5. Proton jump frequencies estimated by CTB-MD in (a) the rotation path and (b) the hopping path. The open circles denote the counted jump frequencies during the MD simulations. The open triangles denote the jump frequencies estimated within the TST but beyond the harmonic approximation. The dashed lines are the estimated proton jump frequencies under the classical h-TST.

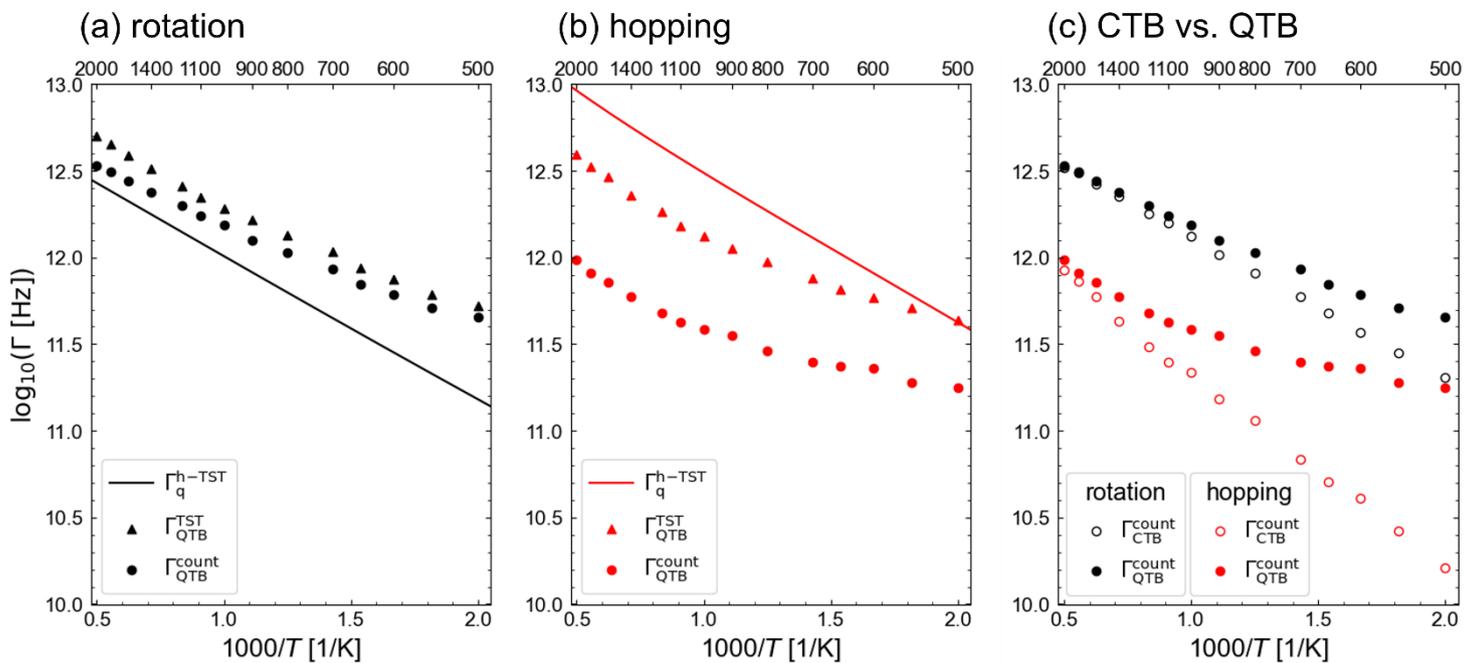

FIG. 6. Proton jump frequencies estimated by the QTB-MD in (a) the rotation path and (b) the hopping path. The solid circles denote the counted jump frequencies during the MD simulations. The solid triangles denote the jump frequencies within the TST but beyond the harmonic approximation. The solid lines are the estimated proton jump frequencies under the quantum h-TST. (c) Comparison in the counted proton jump frequency between the CTB-MD (open symbols) and the QTB-MD (solid symbols).

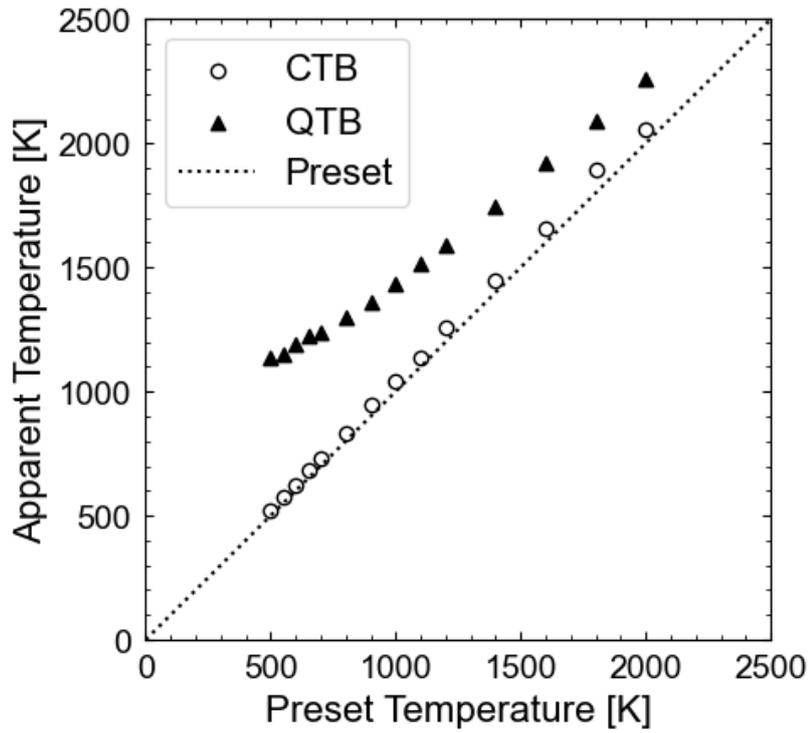

FIG. 7. The apparent temperature of the proton as a function of preset temperature. The open circles and solid triangles denote the apparent temperature in the CTB-MD and the QTB-MD, respectively. The dotted line denotes the preset temperature.

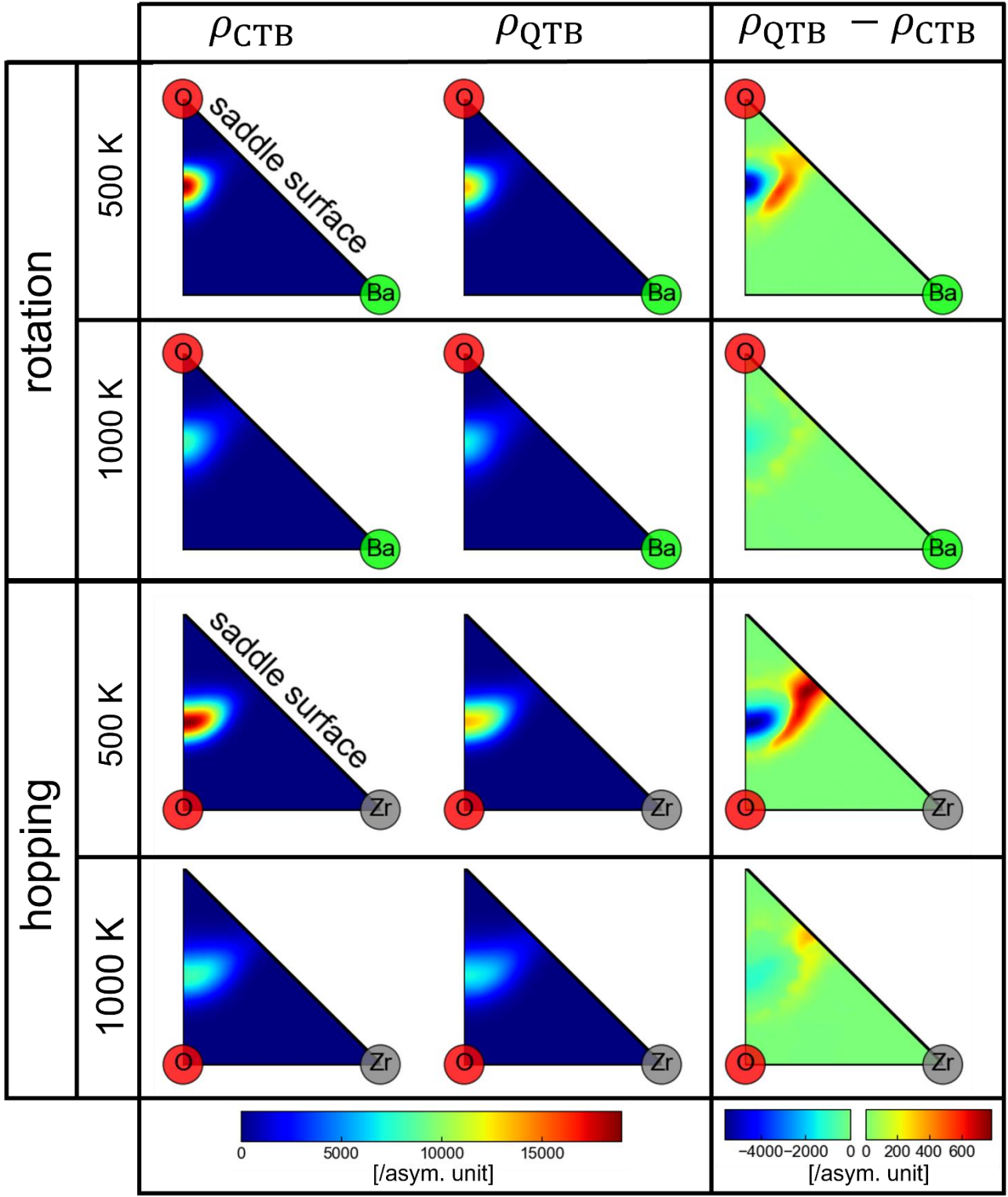

FIG. 8. The left two columns show the probability densities of the proton at 500 and 1000 K obtained from the CTB-MD and the QTB-MD denoted by $\rho_{\text{CTB}}$ and $\rho_{\text{QTB}}$, respectively. The right column shows the difference in the probability density between the CTB-MD and the QTB-MD, $\rho_{\text{QTB}} - \rho_{\text{CTB}}$. The probability densities are shown on the cross-sectional planes including the rotation and hopping paths. The green, grey and red circles denote Ba, Zr and O ions, respectively. The probability density is normalized in the asymmetric unit (Fig. 1), i.e., $\int_{\text{assym. unit}} \rho \, dx = 1$.

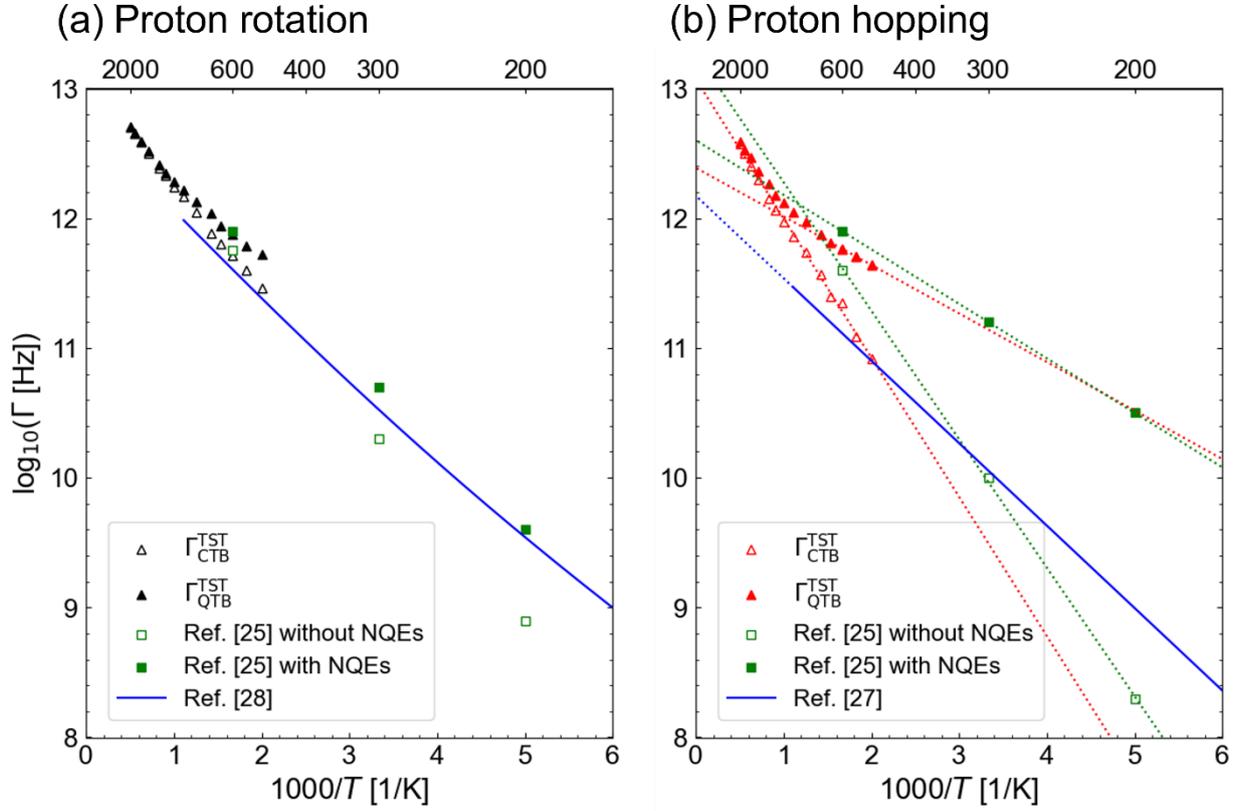

FIG. 9. The proton jump frequencies of (a) proton rotation and (b) proton hopping estimated in the present study and in the previous theoretical studies by Zhang et al. [25] and Geneste [27,28]. The open and solid triangles denote the $\Gamma^{\mathrm{TST}}$ defined by Eq. (22) in the CTB-MD and the QTB-MD. respectively. The green symbols denote those estimated by Zhang et al. using the PIMD, where the open and solid symbols correspond to those without and with the NQEs, respectively. The blue lines denote those estimated by Geneste assuming the adiabatic and non-adiabatic proton tunneling.

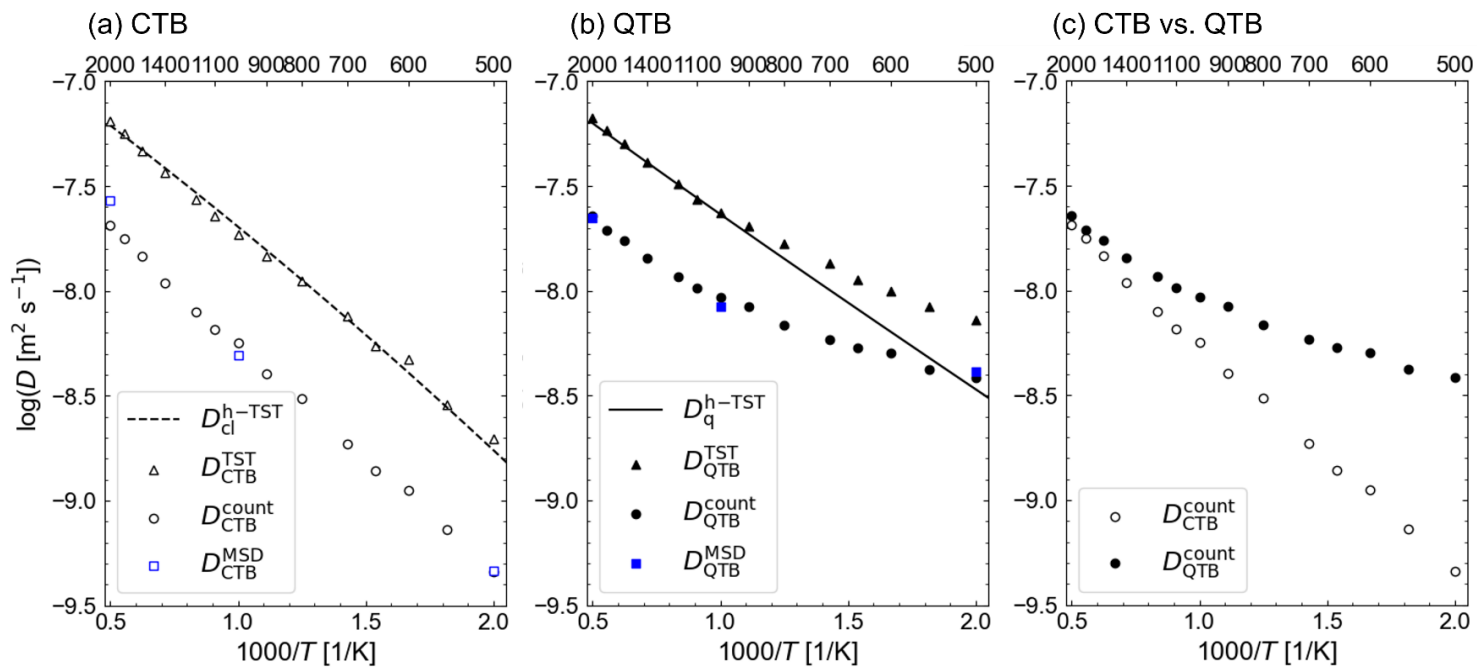

FIG. 10. Diffusion coefficients estimated by (a) the CTB-MD and (b) the QTB-MD. The triangles and circles denote the diffusion coefficients estimated within or beyond the TST, respectively. The diffusion coefficients under the h-TST based on the classical and quantum statistics are also shown by the dashed and solid lines, respectively. The blue squares denote those estimated from the mean square displacements. (c) Comparison between the diffusion coefficients estimated by the CTB-MD and the QTB-MD, shown by the open and solid symbols, respectively.

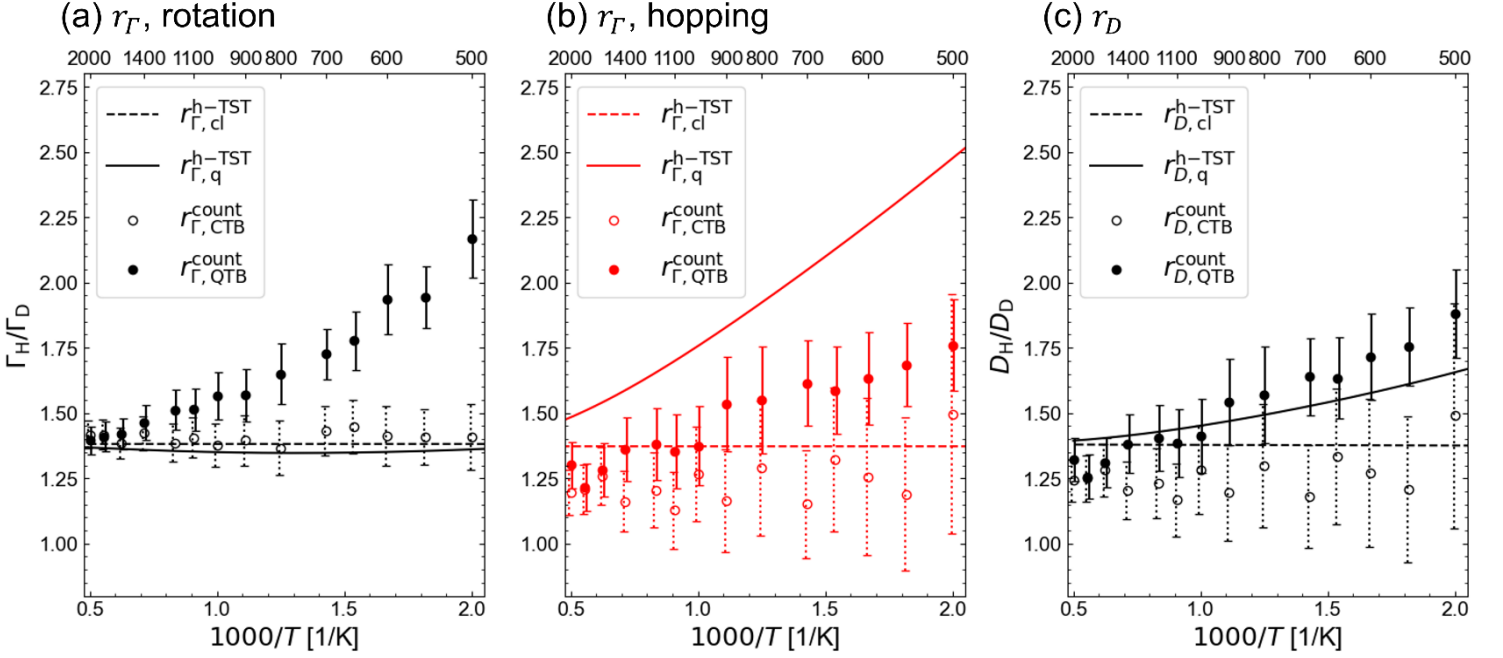

FIG. 11. H/D ratios of the estimated jump frequencies in (a) the rotation path and (b) the hopping path, and (c) those of diffusion coefficients. The open and solid symbols correspond to those estimated from the CTB-MD and the QTB-MD, respectively. The dashed and solid lines denote those within the classical and quantum h-TST, respectively. The error bars were estimated according to the error propagation formula based on twice the standard deviation of the counted jump number of protons or deuterons.

# *Supplementary Materials*

## Nuclear Quantum Effects on Proton Diffusivity in Perovskite Oxides

Shunya Yamada, Kansei Kanayama, and Kazuaki Toyoura*

*Department of Materials Science and Engineering, Kyoto University, Kyoto 606-8501, Japan*

* toyoura.kazuaki.5r@kyoto-u.ac.jp

In the present study, the proton jump frequencies in $BaZrO_3$ were evaluated by the MD simulations with the on-the-fly machine learning force field (MLFF) implemented in the VASP code. The difference in the estimated proton jump frequency between the constructed MLFFs and the first-principles calculations was evaluated to examine the validity of the MLFFs. Figure S1 shows the comparison between the counted proton jump frequencies estimated from the MLFF-MD and the first-principles MD (FPMD) simulations with the classical thermal bath (CTB). The simulation time was set to 2 ns for the MLFF-MD and 0.1 ns for the FPMD after the thermal equilibrium process (0.01 ns). The estimated jump frequencies of proton rotation and hopping by the MLFF-MD are in good agreement with those by the FPMD within the margin of error, suggesting that constructed MLFFs give reasonable estimates of proton jump frequencies.

The quantum thermal bath (QTB) has a well-known problem, the zero-point energy leakage (ZPEL), where the zero-point energies of high-frequency modes are partially transferred to low-frequency modes. In the present system, there is a concern that the large ZPEs of the OH stretching and bending modes are transferred to the host lattice. As a result, the proton and the host lattice become colder and hotter than expected from the preset temperature. If the hotter host lattice accelerates the proton jump significantly, the non-linear behavior of proton jump frequencies in the QTB-MD (Fig. 9) with upward deviation from those in the CTB-MD is physically meaningless. Therefore, we checked the effective temperatures of the host lattice and the proton by converting the mean kinetic energies to the classical temperatures (called *converted temperatures*, hereafter), which are shown in Fig. S2. For comparison, the converted temperatures estimated from the eigenfrequencies under the quantum harmonic approximation are also shown by the broken lines in the figure. It is clear from the figure that the converted temperatures of the host lattice and the proton are higher and lower than expected from the eigenfrequencies, respectively. However, the difference decreases with increasing temperature, and the overheating of the host lattice is as small

# Supplementary Materials

as 40 K even at the lowest temperature, 500 K. Figures S3(a) and (b) show the Arrhenius plots using the effective host lattice temperature (blue solid triangles), which slightly deviates downwards from the plots using the preset temperature. Taking the lowered effective temperature of the migrating proton by the ZPEL into consideration, the blue solid triangles correspond to the lower limits of the proton jump frequencies. In both proton rotation and hopping cases, the blue solid triangles still deviate upwards from the jump frequencies in the CTB-MD, meaning that the non-linear behavior in the QTB-MD corresponds to the NQEs on the proton jump frequencies.

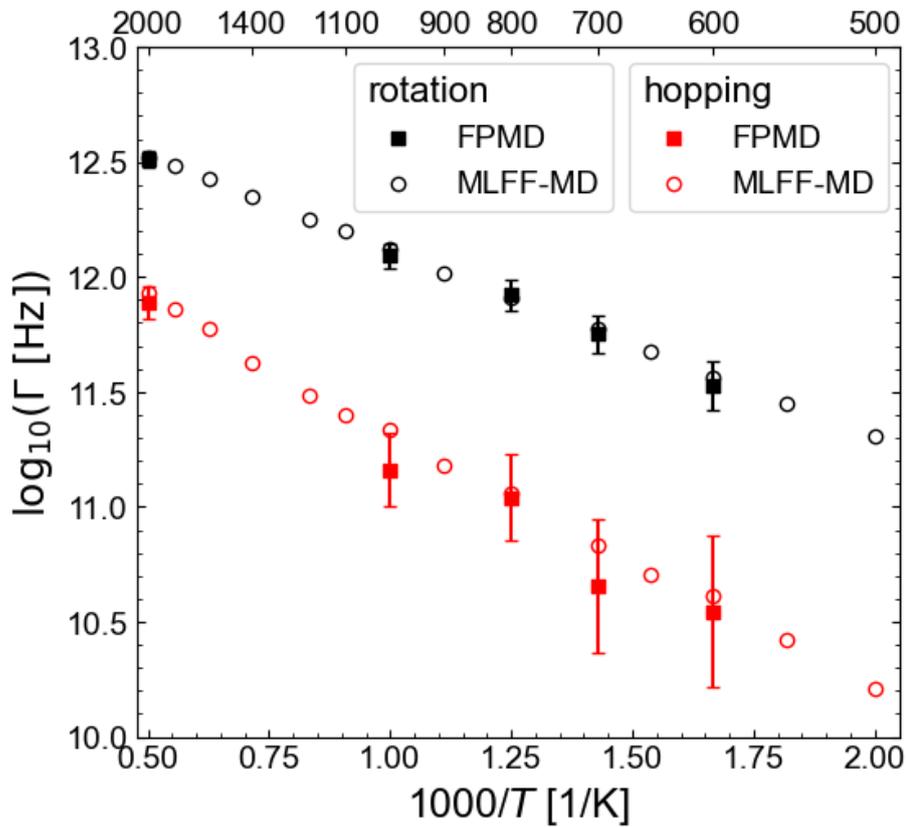

FIG. S1. Comparison between the counted proton jump frequencies estimated from the MLFF-MD (open circle markers) and the FPMD (solid square markers) with the CTB. The black and red symbols denote those in the rotation and hopping paths, respectively. The error bars in the FPMD are $2\sqrt{N_{\text{jump}}}$ corresponding to twice the standard deviation of the Poisson distribution, where $N_{\text{jump}}$ is the counted number of proton jumps.

# *Supplementary Materials*

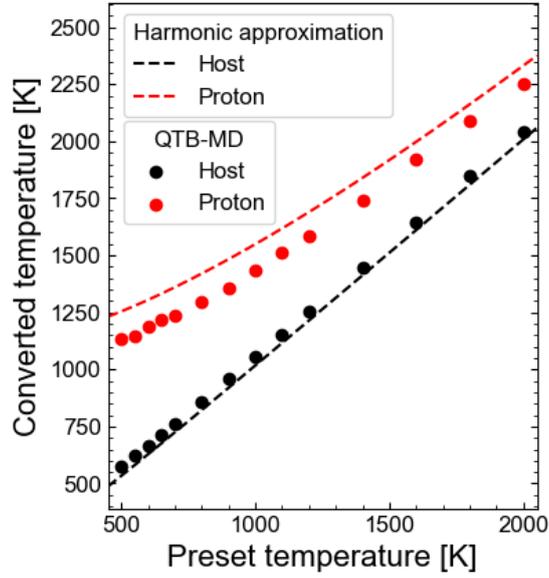

FIG. S2. Classical temperatures of the host lattice (black solid circles) and the proton (red solid circles) converted from the mean kinetic energies in the QTB-MD simulations as a function of preset temperature. The broken lines denote the classical temperatures of the host lattice and the proton, respectively, estimated from the eigenfrequencies of lattice vibration under the quantum harmonic approximation.

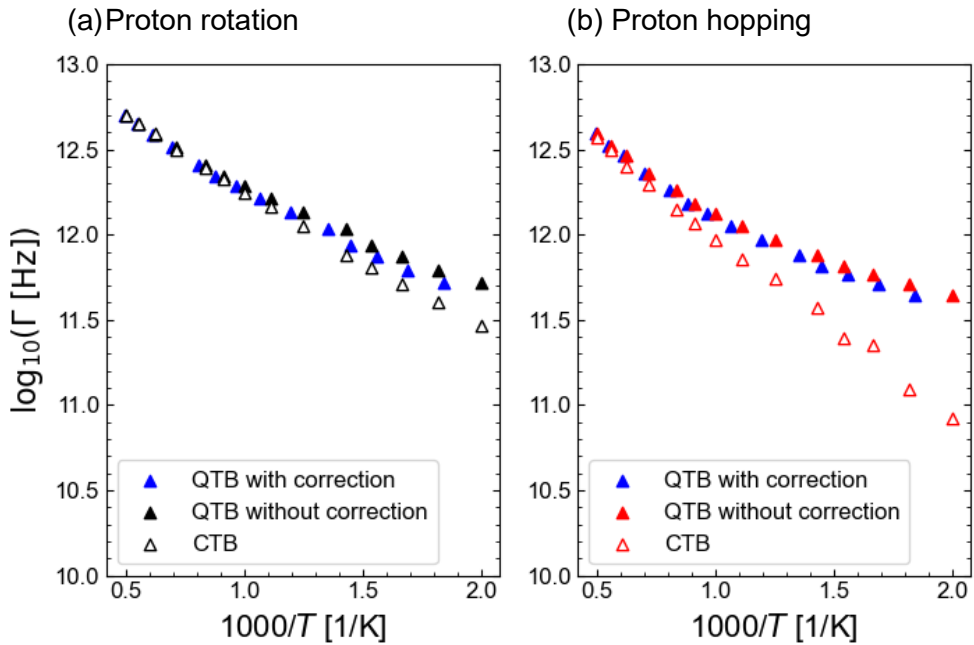

FIG. S3. Arrhenius plots of $\Gamma_{\text{QTB}}^{\text{TST}}$ using the effective host lattice temperature (blue solid triangles) in (a) proton rotation and (b) proton hopping, respectively. The $\Gamma_{\text{QTB}}^{\text{TST}}$ and $\Gamma_{\text{CTB}}^{\text{TST}}$ vs. the preset temperature are also shown for comparison.